# Giant thermal hysteresis in Verwey transition of single domain Fe₃O₄ nanoparticles

Taehun Kim[1,2,*], Sumin Lim[3,*], Jaeyoung Hong[4,5,*], Soon Gu Kwon[4,5,*], Jun Okamoto[6], Zhi Ying Chen[7], Jaehong Jeong[1,2], Soonmin Kang[1,2], Jonathan C. Leiner[1,2], Jung Tae Lim[8], Chul Sung Kim[8], Di Jing Huang[6,7], Taeghwan Hyeon[4,5], Soonchil Lee[3], and Je-Geun Park[1,2,#]

[1]Center for Correlated Electron Systems, Institute for Basic Science, Seoul 08826, Korea

[2]Department of Physics & Astronomy, Seoul National University, Seoul 08826, Korea

[3]Department of Physics, Korea Advanced Institute of Science and Technology, Daejeon 34141, Korea

[4]Center for Nanoparticle Research, Institute for Basic Science, Seoul 08826, Korea

[5]School of Chemical and Biological Engineering, Seoul National University, Seoul 08826, Korea

[6]National Synchrotron Radiation Research Center, Hsinchu 30076, Taiwan

[7]Department of Physics, National Tsing Hua University, Hsinchu 30013, Taiwan

[8]Department of Physics, Kookmin University, Seoul 02703, Korea

* Authors with equal contribution

# Corresponding author: jgpark10@snu.ac.kr

## Abstract

Most interesting phenomena of condensed matter physics originate from interactions among different degrees of freedom, making it a very intriguing yet challenging question how certain ground states emerge from only a limited number of atoms in assembly. This is especially the case for strongly correlated electron systems with overwhelming complexity. The Verwey transition of Fe₃O₄ is a classic example of this category, of which the origin is still elusive 80 years after the first report. Here we report, for the first time, that the Verwey transition of Fe₃O₄ nanoparticles exhibits size-dependent thermal hysteresis in magnetization, $^{57}$Fe NMR, and XRD measurements. The hysteresis width passes a maximum of 11 K when the size is 120 nm while dropping to only 1 K for the bulk sample. This behavior is very similar to that of magnetic coercivity and the critical sizes of the hysteresis and the magnetic single domain are identical. We interpret it as a manifestation of charge ordering and spin ordering correlation in a single domain. This work paves a new way of undertaking researches in the vibrant field of strongly correlated electron physics combined with nanoscience.

Strongly correlated electron systems are a group of materials, where electron correlations determine a delicate balance among the competing phases of the systems[1-3]. Naturally, the challenge becomes ever so mounting when one tries to handle a case where there are correlations involving not just one but all four degrees of freedom of solid: charge, orbital, lattice, and spin. This is exactly the case for the Verwey transition of $Fe_3O_4$, where the electronic degree of freedom (charge) is tightly coupled to all three others[4-8]. As the temperature becomes lower than the Verwey transition temperature, $T_V$, a number of events take place simultaneously: charge ordering of $Fe^{3+}$ and $Fe^{2+}$ cations, the lattice symmetry change from cubic (Fd$\bar{3}$m) to monoclinic ($C_c$), metal-insulator transition, sudden drop in magnetic susceptibility, and the magnetic easy axis change from (111) to (100). Resonant X-ray scattering technique also showed a hint of orbital ordering[9]. Due to such complexity of these correlated phenomena, the origin of the Verwey transition still remains unresolved even after about 80 year since its first report[4].

The central question of strongly correlated electron systems is how the strong electron correlations emerge out of the assembly of individual uncorrelated atoms. To be specific to the Verwey transition of $Fe_3O_4$, it will be a very interesting study to investigate the size dependence of the transition. Given the nature of the complex interactions, it may well offer a new window of opportunities looking into the intimate details of the Verwey transition. The recent advance in nanoscience has brought new methodology to solid state physics for studying the emergent properties of a limited number of atoms in assembly. Through solution-based synthesis, colloidal nanoparticles (NPs) can be prepared with a precise size control in the range from sub-nanometer to tens of nanometers, which amounts to the number of atoms ranging from $10^1$ to $10^{20}$. In terms of electronic energy structure (molecular orbital to band) and symmetry of atomic arrangement (point group to space group), transition from molecule to solid takes place at around $10^2$-$10^5$ atoms[10]. Armed with the latest huge developments in the nanochemistry field, one can open a new, exciting opportunity of monitoring the emergence of electron correlations at the borderline between molecules and solids. For example, the critical pressure of the solid-solid phase transition for CdSe NPs significantly increases as their sizes are reduced to less than few nanometers[11], whereas the melting temperatures of CdS and metal NPs show an opposite trend[12-14]. Nevertheless, studies are relatively rare on more complex transitions emerging from electron correlations for colloidal NPs.

Previously, we developed a fine synthetic method to prepare stoichiometric $Fe_3O_4$ NPs with various sizes to observe that $T_V$ becomes size dependent for the NPs smaller than 40 nm[15]. Here, we show another, yet hitherto unreported feature of the Verwey transition. By probing all the four degrees of freedom using independent techniques with superconducting quantum interference device (SQUID) magnetometer, cryogenic X-ray diffraction (XRD), $^{57}Fe$ nuclear magnetic resonance (NMR), and resonant inelastic X-ray scattering (RIXS), we reveal that the Verwey transition of stoichiometric $Fe_3O_4$ NPs exhibits large size-dependent thermal hysteresis and that the hysteresis width is correlated with the critical size of the magnetic single domain, a feature is not discovered yet.

**Results**

The Verwey transition is extremely sensitive to oxidation such that it is substantially suppressed when off-stoichiometry parameter δ, defined as $Fe_{3(1-δ)}O_4$, is larger than only 1%[16]. Therefore, we conducted preparation and storage of our $Fe_3O_4$ NPs with extreme care to exclude any possible cause of off-stoichiometry[15]. $Fe_3O_4$ NPs with the sizes from 7 to 390 nm are prepared under 4 wt% $CO/CO_2$ atmosphere in order to set the redox equilibrium of $Fe^{3+}$ and $Fe^{2+}$ at the ratio of 2:1 in inverse spinel structure [$(Fe^{3+}_8)_{tet}(Fe^{2+}_8, Fe^{3+}_8)_{oct}O_{32}$]. As a bulk standard sample, 7 μm particles are also prepared via thermal annealing of iron oxide powder under reducing atmosphere. Throughout the experiments, NPs are kept under inert atmosphere being strictly isolated from any oxidation source. Size distribution data from transmission electron microscopy (TEM) show that the NPs smaller than 100 nm have narrow size distribution with the relative standard deviation (r.s.d.) of ~10% or lesser, while the larger ones have broader distribution with r.s.d. of 17-29% (Figs. S1 and S2 in Supplementary Information). To check the $Fe^{2+}/Fe^{3+}$ ratio, we performed the Mössbauer measurements on 7 μm and 42 nm samples. The results are summarized in Fig. S7 and Table S1 in SI. Obtained $Fe^{2+}/Fe^{3+}$ ratio for 7 μm is 1/1.9994 and that for 42 nm is 1/1.9985, which corresponds to the off-stoichiometry parameter $\delta$ = 0.0002 for 7 μm and $\delta$ = 0.0005 for 42 nm. Based on these results, we confirmed that both samples are high-stoichiometric within the error of $\delta$ = ±0.0003.

The magnetization data of our NPs show a clear correlation between thermal hysteresis and the particle size in nanometer scale (Fig. 1). Bulk standard sample (7 μm) exhibits a characteristic drop in magnetization at the known value of $T_V$ with a thermal hysteresis of 1 K, consistent with those reported in literature[17]. For the NPs, on the other hand, as the size decreases from 390 to 120 nm, the width of the thermal hysteresis increases from 2 to 11 K before falling down until the Verwey transition itself disappears below 10 nm[15]. When the heating/cooling rate varies from 0.1 to 10 K min$^{-1}$, the value of $T_V$ as well as the hysteresis width do not show any significant change within error bars (see Fig. S3 in SI). In controlled oxidation experiment done on 7 μm and 16 nm samples, the thermal hysteresis width of 16 nm sample after oxidation significantly change from 4.6 K to 0.7 K and that of 7 μm sample decrease from 1 K to 0.3 K (see Fig. S4 in SI). Therefore, we can rule out any form of sample degradation from possible explanation, especially in the form of oxidation. It implies that the size-dependent thermal hysteresis observed from the NPs is most likely to be an intrinsic effect.

Our subsequent $^{57}Fe$ NMR measurements tell a similar story too. The NMR signal measured above $T_V$ (130 K) shows a very narrow single peak for the 7 μm sample (Fig. 2a). This peak progressively gets broadened and shifted to the lower frequencies with reducing size, indicating inhomogeneous spin dynamics in the NPs due to the high surface-to-volume ratio. When measured below $T_V$ (110 K), the single peak of the 7 μm sample gets split into several ones, which is a clear sign of a new magnetic phase with quite different internal magnetic field distribution[18]. This general tendency is consistently seen for the NP samples. It is in stark contrast with the almost flat signals measured on $\gamma\text{-}Fe_2O_3$ (maghemite) under the identical condition, which is another evidence of our experimental observations being intrinsic effect of NPs on the Verwey transition, not due to oxidation. Yet a more drastic observation is made on the temperature dependent NMR data (Fig. 2b). For the 7 μm sample, the single peak splits into weakened multiple peaks and reappears at 120-121 K upon cooling and heating with almost the same peak widths. On the other hand, the NMR data collected on the 42 nm sample display a significant thermal hysteresis; the single peak is split at 108.5 K in cooling and recovered at 119.5 K in heating with a thermal hysteresis as large as 11 K. The size-dependent hysteresis of the NMR data becomes much clearer when the broadness of the NMR spectra are plotted against the temperature (Fig. 2c). Note that the spectra broadness present in Fig. 2c is defined at quarter maximum of the strongest NMR peak at each temperature in the same frequency region. The hysteresis decreases with the size of the NPs smaller than 77 nm, as we observed from the magnetization data in Fig. 1b.

The thermal hysteresis is also registered in the temperature-dependent XRD data (Fig. 3a-3b). Upon the Verwey transition, $Fe_3O_4$ (440) peak becomes significantly broadened due to the change in the lattice structure. The transition temperatures of the 42 nm sample measured in cooling and heating procedures differ by 7 K, in agreement with both the magnetization and the NMR data discussed above. We also undertook RIXS experiments at near Fe $L_3$-edge with σ polarization on the samples at the temperatures above (300 K) and below (40 K) $T_V$. As shown in Fig. 3c, there is a clear peak centered at 0.2 eV for the 7 μm sample at 300 K. This peak is assigned to a polaron excitation in a recent $Fe_3O_4$ single crystal study[19]. With reducing the size, this peak gets progressively suppressed and almost disappears for the 7 nm sample. For comparison, our data collected on $\gamma$-$Fe_2O_3$ do not show the peak at all. When we compare the RIXS data collected at 40 and 300 K, there is no visible temperature dependence as one can see in Fig. 3d (see Fig. S5 in SI for multiple peak fittings to obtain the peak area): a recent RIXS measurement made on single crystal $Fe_3O_4$ showed that the polaron peak persists well into the high temperature phase with very small temperature dependence up to 550 K[19]. Similarly, the peak area does not show much of temperature dependence either. It is to be noted that the measurements were taken with the energy resolution of 100 meV and the spectra with π polarization incident beam also show the same excitation.

With all the data taken together, the size dependence of the thermal hysteresis becomes much clearer as shown in Fig. 4a and 4b. The hysteresis width, $\Delta T_V = T_V(heating) - T_V(cooling)$, shows clear size dependence with the critical size of 120 nm. For the sizes larger than or equal to 120 nm, $\Delta T_V$ is proportional to $D^{-1}$ where $D$ is the mean size of the NPs. On the other hand, it shows $-D^{-3/2}$ behaviour for $D < 120$ nm. We note that this size dependence of $\Delta T_V$ is very similar to the size-dependent magnetic property of the NPs[20]. As shown in Fig. 4c, the ratio of the remnant and saturated magnetization, $M_r/M_s$, from isothermal magnetization at 20 K gradually increases with reducing size, before flattening off at a value of 0.4 far below 120 nm (see Fig. S6 in SI for the raw data). At 300 K, this parameter shows a peak at the same size. This observation is consistent with the theoretical prediction for $Fe_3O_4$ that the value of $M_r/M_s$ for a single domain is 0.5[21], indicating that the NPs become magnetic single domain below the critical size of 120 nm. Transition from multi- to single domain is also corroborated by the size-dependent coercivity data, $H_c(D)$, in Fig. 4d that show the maximum at around 120 nm. Furthermore, the fitting curves from the $\Delta T_V$ data is remarkably similar with the coercivity of the NPs. Based on this relationship between the size dependence of $\Delta T_V$ and magnetism, we conclude that the Verwey transition and magnetic domain share the same critical size.

**Discussion**

Let us now turn our attention to the origin of the observations we made above: i) the size-dependent thermal hysteresis of the Verwey transition, $\Delta T_V(D)$, and ii) the coincidence of the critical size of $\Delta T_V(D)$ and magnetic domain. In general, the width of the hysteresis reflects the kinetic energy barrier for the phase transition. The hysteresis of 11 K (0.95 meV) at its maximum gives a rough estimation of energy correlated with charge ordering which is considered to drive the Verwey transition[8]. As we discussed above, the Verwey transition has the composite nature of intercoupling between the four degrees of freedom in solid: lattice, orbital, spin, and charge. Among them, the energy scale of spin degree of freedom (magnetism) is the smallest and about a few meV, being closest to that of $\Delta T_V$, while the others usually have larger characteristic energy scales, often of a few eV. According to the previous literature, iron oxide NPs with the size of 160 nm or smaller have a single lattice domain[22,23]. Therefore, we can safely assume that our NPs form in a single domain of all four degrees of freedom below 120 nm, and the size of magnetic single domain has dominant effect on the kinetics of the Verwey transition.

Study on the relationship between the kinetics of phase transition and the size of colloidal NPs dates back to two decades ago[11,24]. In 1997, it was reported that the high pressure-induced solid-solid phase transition of CdS NPs show size-dependent hysteresis with the kinetic energy barrier rapidly increasing with the size of the NPs from 1.5 up to 4.3 nm[24]. Apparently, this observation has strong similarity with our result that the NPs have a higher energy barrier for the phase transition compared to bulk solid and that the hysteresis width passes a maximum as the size gets reduced. For CdS NPs, it was suggested that the size-dependent hysteresis is related with the single nucleation event within a nanoparticle upon the phase transition. As long as the size of a nanoparticle is smaller than the critical size of the nucleus of another phase, a phase transition is completed by a single nucleation and the energy barrier is linearly proportional to its volume. When the particle size is large enough ($10^2$-$10^3$ nm), multiple nucleation and growth take place within the volume, which lowers the energy barrier with respect to that of the single nucleation (see Scheme S1 in SI for the detail of the model). However, this model does not apply to our data: $\Delta T_V(D)$ is not proportional to $D^3$ but $-D^{-3/2}$ when $D <$ 120 nm, and the crossover from $-D^{-3/2}$ to $D^{-1}$ behaviour is abrupt and discontinuous, rather than gradual change from single nucleation to multiple nucleation.

The significant resemblance of $\Delta T_V(D)$ and $H_c(D)$ in Figs. 4b and 4d suggests that the coupling between charge and spin degrees of freedom is manifested in the form of metastability in the Verwey transition. For a uniaxial magnetic single domain, $H_c$ is proportional to $-D^{-3/2}$ due to a balance between the energy barrier for coherent spin rotation and thermal energy[20]. In a magnetic multi-domain regime, domain wall nucleation is at work to lower the energy barrier for spin rotation. In other words, $H_c$ reflects the kinetic stability of magnetization reversal. And, in the Verwey transition, $\Delta T_V$ has the same meaning for transition between charge ordering and disordering. At this point, the mechanism underlying correlation of charge and spin ordering is not understood yet. Given the nature of strong correlations between the spin and charge degrees of freedom, one can understand this similarity in the way that when domains of the charge channel get modified upon reducing size naturally the corresponding domain of the spin channel will follow the suit through the anticipated strong coupling between the spin and charge degrees of freedom. Therefore, it seems to us natural to observe such a strong similarity between $\Delta T_V(D)$ and $H_c(D)$. Interestingly, an increased thermal hysteresis was also observed in thin film when $Fe_3O_4$ thin films were prepared under optimized oxygen partial pressure to match the stoichiometry[25]. In this case, they cited that this thermal hysteresis is not directly related to the average domain size of film, but instead they suggest that anti-phase boundaries may affect the thermal hysteresis.

To summarize, we observed that the Verwey transition of stoichiometric $Fe_3O_4$ NPs shows a very large size-dependent thermal hysteresis. For our bulk standard sample, the hysteresis width is only 1 K. In nanometer scale, however, the hysteresis gets significantly enhanced reaching the maximum of 11 K for 120 nm sample. For the NPs smaller than 120 nm, the hysteresis width is well fitted with $-D^{-3/2}$ curve, similar to the size dependence of the coercivity. Also, the size of the maximum hysteresis coincides with the critical size of magnetic single domain. Overall, $\Delta T_V(D)$ and $H_c(D)$ show significant resemblance. We think that this close relationship suggests a much tighter coupling than have been thought of the charge and spin degrees of freedom in $Fe_3O_4$.

## Methods

**Synthesis of stoichiometric $Fe_3O_4$ nanoparticles.** For the precise stoichiometry control, the whole procedures were performed under slightly reducing atmosphere by using a standard Schlenk technique under a gas flow of 4 wt% $CO/CO_2$[15]. All of the solvents were deaerated and stored in a glove box before use. In a typical synthesis, a mixture of $Fe(acac)_3$ and oleic acid in benzyl ether was degassed under vacuum for 1 hr. The mixture was heated to 290 °C at a rate of 20 K min$^{-1}$ and then kept at the same temperature with vigorous stirring for 30 min to complete the reaction. $Fe_3O_4$ NPs were separated by adding 200 ml of toluene and 500 ml of ethyl alcohol to the crude mixture under an inert atmosphere. After washing, NPs were collected by centrifugation and stored in a glovebox in the form of dried powder. The size of NPs was controlled by varying the relative amount of reactants. For the NPs smaller than 100 nm, a mixture of 33.9 g of oleic acid in 312 g of benzyl ether was used with varying amount of Fe(acac)$_3$ from 24.5 g for 77 nm NPs to 5.33 g for 7 nm ones. For the larger NPs, 3.39 g of oleic acid in 31.2 g of benzyl ether was used with 2.34 g and 1.92 g of Fe(acac)$_3$ for 390 and 120 nm NPs, respectively. To prepare bulk standard Fe$_3$O$_4$ (7 μm particles), magnetite powder from a commercial source was annealed at 1300 °C for 24 hours under CO gas flow to obtain the exact stoichiometry. To see the effect of off-stoichiometry, 7 μm sample was deliberately oxidized under air at 200 °C for six days and the NP samples in ambient air at room temperature for four days.

**Characterizations.** Size distribution and morphology of Fe$_3$O$_4$ NPs were analyzed by using a JEOL JEM-2010 transmission electron microscope operating at 200 kV. Magnetic susceptibility was measured with a Quantum Design SQUID magnetometer MPMS 5XL. To scan the thermal hysteresis, a sample was cooled down to 20 K at zero field first. Then measurement was carried out at 100 Oe from 20 to 200 K and from 200 to 20 K in a row. Note that a conventional zero field cooling and field cooling curve is not same with our experiment, because our purpose of measurements is to see the thermal hysteresis of Verwey transition. The value of $T_V$ was obtained from Gaussian fit of d$M$/d$T$ curves. To define the values of the remanent and saturated magnetization, we also carried out isothermal magnetization measurements from −6000 to 6000 Oe at 20 and 300 K, respectively. Cryogenic XRD measurement was conducted using a Bruker D8 Discover System with Oxford cryosystems. NMR spectra were taken using a home-made solid state NMR spectroscopy instrument equipped with a cryostat. We swept the frequency from 67.5 to 71.5 MHz under zero field at different temperatures in the range from 90 to 130 K. For both XRD and NMR data, $T_V$ was calculated by fitting derivative of the peak width with respect to the temperature with a Gaussian curve. Measurement time for each temperature, including both temperature stabilization and scan time, was 2-3 min for magnetometer, 50 min for XRD, and 1 hour for NMR, respectively. RIXS experiments were performed at 05A1 beamline of the Taiwan Light Source at the National Synchrotron Radiation Research Center (NSRRC) in Taiwan [19, 26]. We used an AGM-AGS spectrometer and the scattering angle was set at 90 degree. The measurements were conducted at zero field in a high vacuum chamber (~10$^{-9}$ Torr). The incident X-ray energy was defined by X-ray absorption spectrum and fixed at Fe L$_3$ edge minus 4 eV (~706 eV). All samples were prepared in pelletized form with the diameter of 5 mm. We carried out the measurements using both σ and π polarized X-ray at 300 and 40 K. The experimental resolution of energy loss is about 100 meV. The Mössbauer spectra were obtained by using a transmission mode at 295 K. The Mössbauer spectrometer of the electromechanical type consists of a fixed absorber and a moving source in constant-acceleration mode with a $^{57}$Co source of 50 mCi in a rhodium matrix. In addition, Mössbauer spectrometer was calibrated by using an α-Fe foil. The Mössbauer spectra were analyzed by a least-squares fitting procedure and provided the magnetic hyperfine field ($H_{hf}$), Isomer shift (Δ), electric quadrupole splitting ($E_Q$), and relative area the ratio of Fe sites.

**Acknowledgments**

This work at SNU was supported by the Research Center Program of Institute for Basic Science (IBS) in South Korea (Grant No. IBS-R009-G1, IBS-R006-D1, and IBS-R006-Y1). Work at KAIST was supported by the National Research Foundation Grand No. NRF-2015R1A2A1A15055468. Work at the NSRRC was supported by the Ministry of Science and Technology of Taiwan under Grant No. 103-2112-M-213-008-MY3. Work at Kookmin University was supported by the National Research Foundation Grand No. NRF-2017R1A2B2012241.

## Authors Contributions

## Competing financial interests

The authors declare no competing financial interests.

**Figure Captions**

Figure 1. Magnetization curves of Fe$_3$O$_4$ NPs with various sizes and bulk standard (7 μm) in (a) full scale and (b) magnification scale near the transition temperature. The samples are cooled down to 20 K at zero field and the magnetization measured from 20 K to 200 K and from 200 K to 20 K continuously. All measurements are done under an external field of 100 Oe and temperature rate is fixed at 1 K/min around the transition temperature.

Figure 2. (a) $^{57}$Fe NMR spectra of Fe$_3$O$_4$ samples at 130 and 110 K. (b) Temperature-dependent NMR data of 7 μm and 42 nm samples during cooling and heating procedures. Asterisks indicate the Verwey transition. (c) Plots of full width at quarter maximum (FWQM) of the NMR peak vs. the measurement temperature for the samples with various sizes.

Figure 3. (a) The contour plots for XRD data measured during heating and cooling of the 42 nm sample showing temperature evolution of Fe$_3$O$_4$ (440) peak. (b) FWHM of (440) peak in panel a is shown as a function of temperature. (c) RIXS spectra of various Fe$_3$O$_4$ NPs and bulk standard at 300 and 40 K using σ-polarized beam with photon energy at Fe L$_3$ edge minus 4 eV. (e) The peak area of the extracted low energy excitation centered at 200 meV as a function of the size.

Figure 4. (a-b) Size dependence of (a) $T_V$ and (b) $\Delta T_V$ from magnetization and NMR data in Figs. 1 and 2. Coefficient of determination ($R^2$) for fit curves are 0.995 for $-D^{-3/2}$ and 0.991 for $D^{-1}$, respectively. (c-d) Size dependence of (c) $M_r/M_s$ ratio and (d) $H_c$. Fitting curves in panel (d) are identical to those in panel (b).

Figure 1

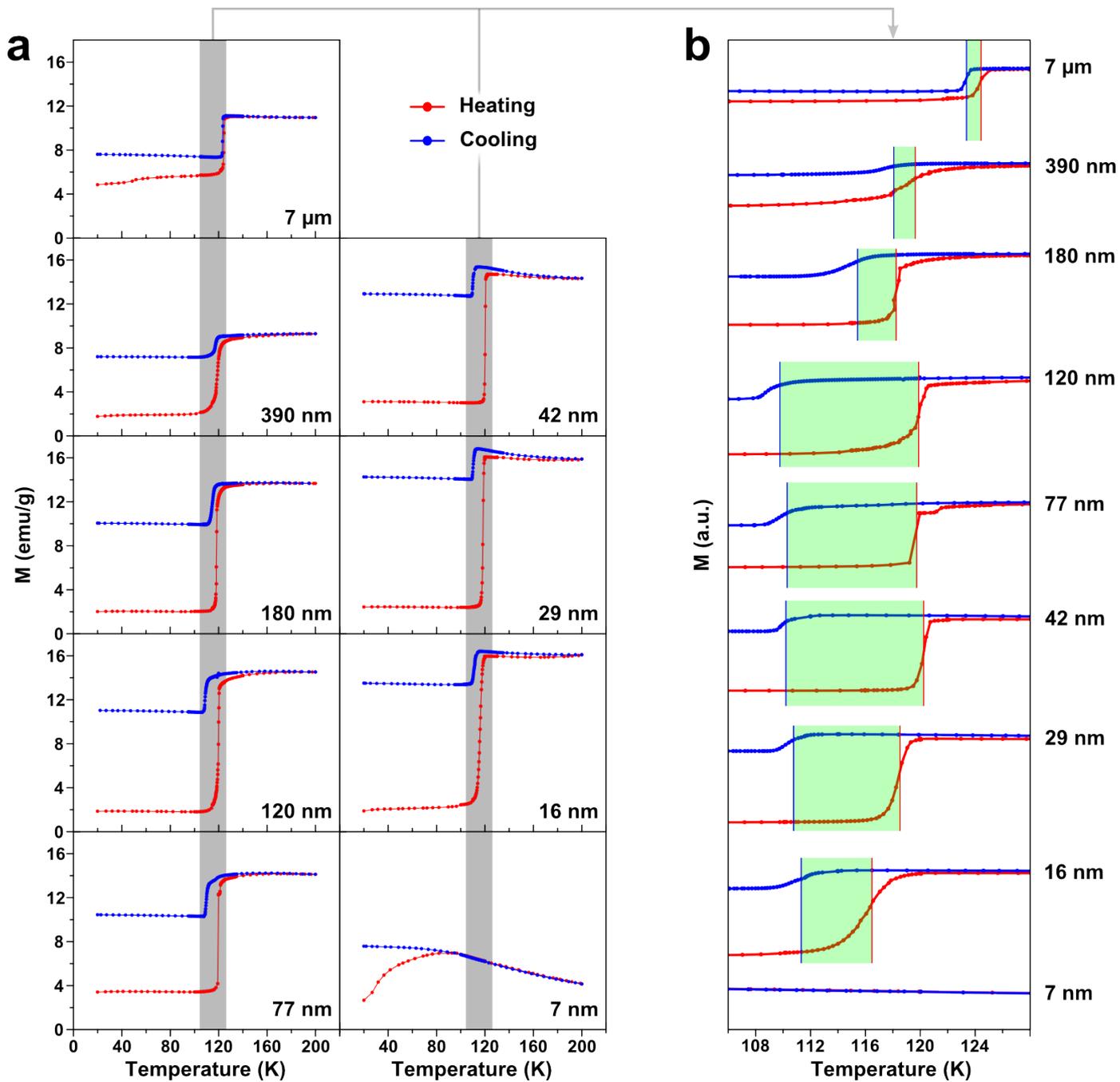

Figure 2

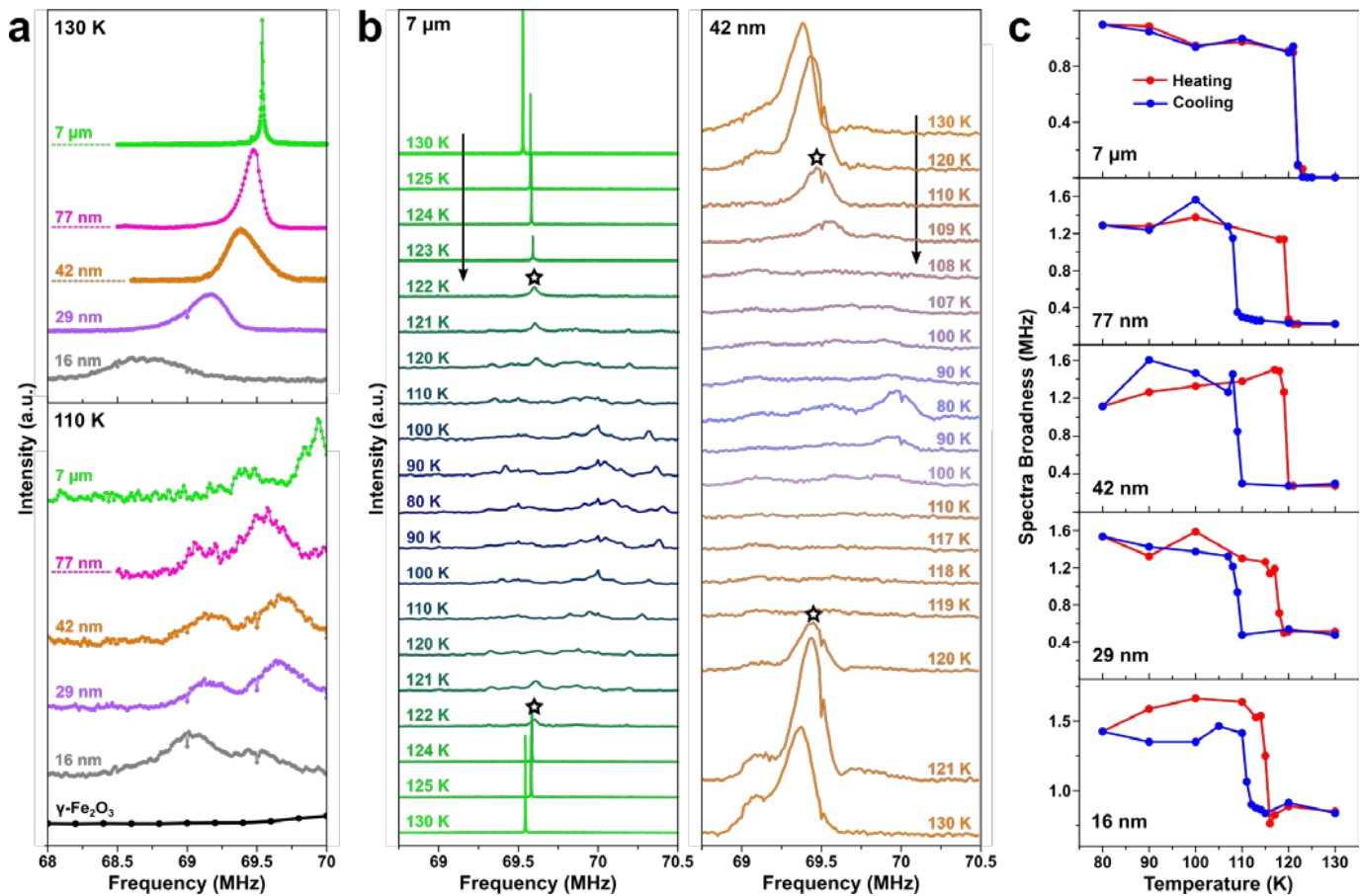

Figure 3

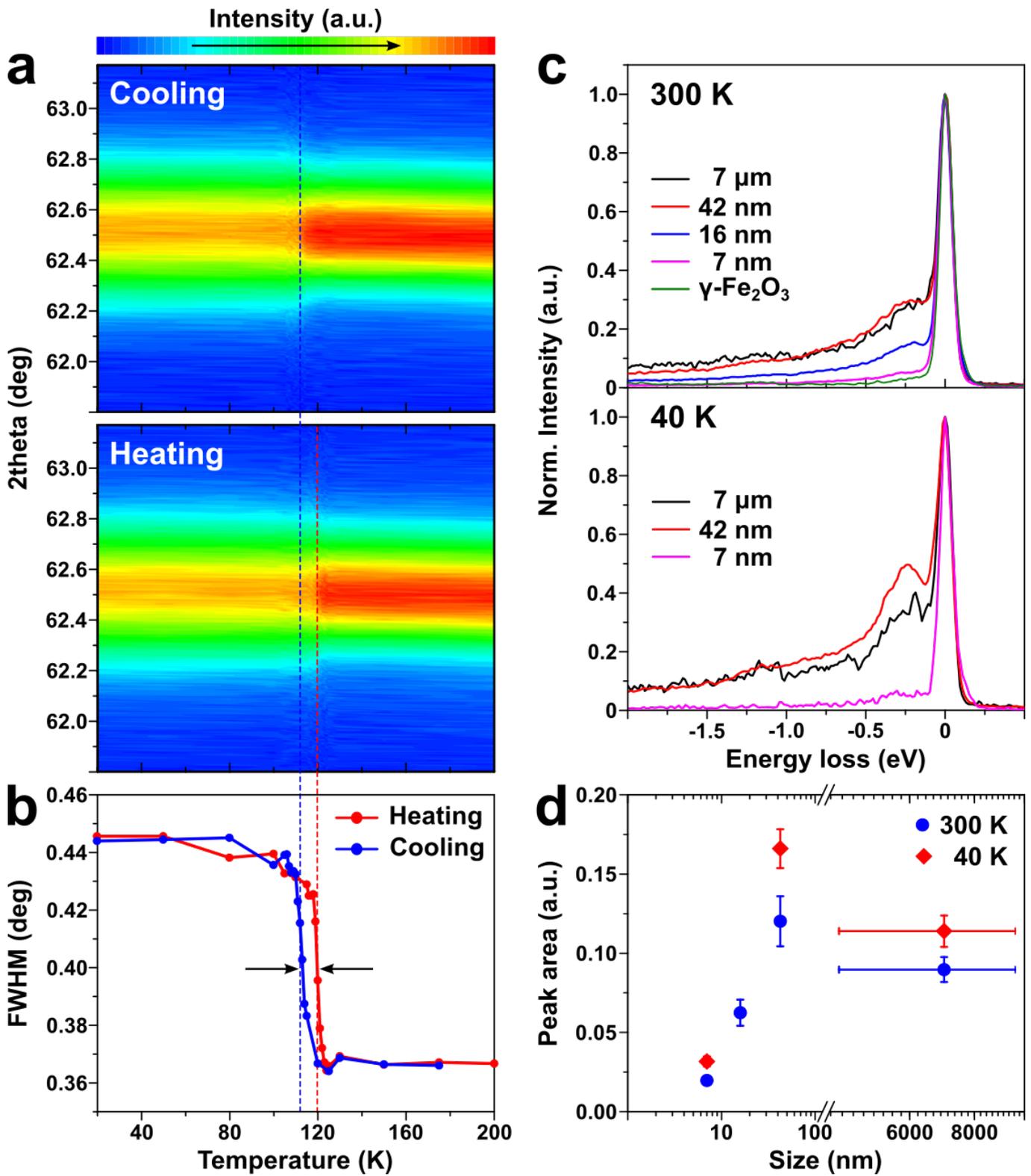

Figure 4

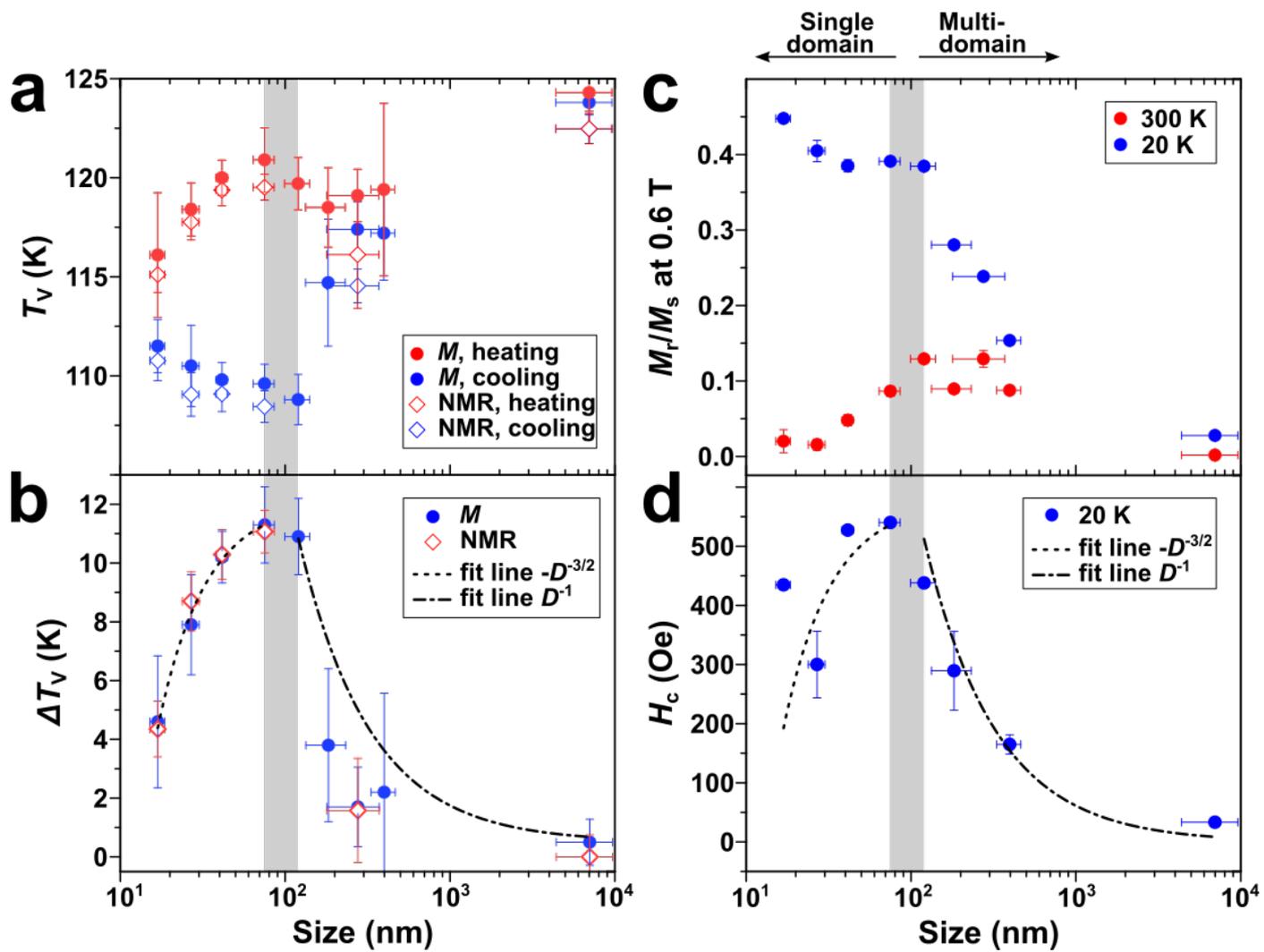